\documentclass[12pt]{article}
\usepackage{graphicx}

\begin{document}

\title{\textbf{TWO-MIRROR\\ SCHWARZSCHILD APLANATS.
BASIC RELATIONS}\thanks{\textit{Astronomy~Letters,~Vol.~31,~No.~2,~2005,
~pp. 129~--139.\qquad \qquad \qquad Translated from Pis'ma v
Astronomicheski$\breve \imath$ Zhurnal, Vol.~31, No.~2, 2005, pp.
143~--153. Original Russian text \copyright \,2005 by Terebizh.}}}

\author{V.~Yu.~Terebizh\thanks{98409 Nauchny, Crimea, Ukraine;
 \,E-mail:\, \textsf{terebizh@crao.crimea.ua}}\\
 \small{\textit{Sternberg Astronomical Institute, Moscow, Russia}}}

\date{\footnotesize{Received July~22, 2004}}

\maketitle

\begin{quote}
\small{\textbf{Abstract}~--- It is shown that the theory of aplanatic
two-mirror telescopes developed by Karl Schwarzschild in 1905 leads to the
unified description both the prefocal and the postfocal systems. The class of
surfaces in the ZEMAX optical program has been properly extended to ascertain
the image quality in exact Schwarzschild aplanats. A comparison of
Schwarzschild aplanats with approximate Ritchey--Chr\'etien and
Gregory--Maksutov aplanatic telescopes reveals a noticeable advantage of the
former at fast focal ratio
of the system.\\
 \copyright \, \textit{2005 MAIK ``Nauka/Interperiodica''.}

\medskip

}
\end{quote}

\newpage

\section*{INTRODUCTION}

In a series of papers written a century ago, Karl Schwarzschild laid the
foundations of the modern aberration theory of optical systems, including
telescopes (see Born and Wolf~1999, Ch.~5; Wilson~1996, 3.2). The last
section in part~II of this series (Schwarzschild~1905) is devoted to
seeking for an aplanatic two-mirror telescope, i.e., a system in which both
spherical aberration and coma were \emph{rigorously} corrected near to an
optical axis. Schwarzschild managed to derive closed analytical formulas
that described the shape of the mirror surfaces in such a telescope. Since
he was primarily interested in fast systems (the photographic emulsions
were then slow), a telescope with a primary mirror of moderate focal
ratio~$f/2.5$ and a concave secondary mirror that brought the total focal
ratio of the telescope to~$f/1.0$ was considered as an example (see Fig.~5
below). Subsequently, Born and Wolf (1999, \S4.10.2) gave a general
formulation of the problem of the simultaneous correction of spherical
aberration and coma in an arbitrary optical system. Grazing-incidence
Schwarzschild telescopes designed for X-ray observations were considered by
Popov~(1980, 1988)\footnote{After the completion of this paper, we learned
about the papers by Lynden-Bell~(2002) and Willstrop and Lynden-Bell~(2003)
devoted to two-mirror aplanats. Having only part~X of the fundamental study
by Schwarzschild~(1905) at their disposal, these authors have repeated in a
different form some of the results contained in part~II of that study.}.

Quite possible, the context in which Schwarzschild discussed the problem and
the example mentioned above have spawned the widely held view that
Schwarzschild's theory is applicable only to prefocal reducing systems (see the
next section for an explanation of the terms). Meanwhile, this theory covers
not only all prefocal systems, including Cassegrain telescopes, but also
postfocal Gregorian systems. Moreover, Schwarzschild's formulas that define the
mirror surfaces in an aplanatic telescope can be brought to a form that is
valid for an arbitrary two-mirror system.

Since the analytical description of the shape of the mirror surfaces in exact
aplanats is rather complex in form, the image quality provided by these
telescopes has remained not cleared up until now. Only the approximations of
the surfaces by conic sections admissible for systems with slow focal ratios
were considered. The expansions that emerge in this case were found by
Schwarzschild~(1905) himself. Subsequently, Chr\'etien~(1922) and
Maksutov~(1932) concretized these expansions for Cassegrain and Gregorian
systems, respectively, which gave rise to the telescopes aplanatic in the third
order of the aberration theory. Most of the modern major instruments belong to
this class. The merits of these systems and the discovery of a fast wide-field
camera by Bernhard Schmidt~(1930) were responsible for the prolonged lack of
interest in Schwarzschild's exact theory.

This situation was explicable as long as the diameter of Schmidt telescopes
corresponded to the needs and technology capabilities of the time. At present,
one can point out several problems of observational astronomy in connection
with which it would be interesting to return to Schwarzschild's theory and
ascertain the image quality achievable with \emph{exact} aplanats. In
particular, one of the most important problems is to carry out deep sky surveys
using telescopes with an effective aperture of 4--7~m and a ${2^\circ
-3^\circ}$ field of view. This will allow to study the evolutionary
cosmological models and to follow the changes in the positions and brightness
of faint objects on a time scale of the order of a day within the entire
celestial sphere. An example of another vital problem involves far-ultraviolet
and X-ray observations from spaceborne platforms, suggesting the use of
telescopes with a minimum number of reflective surfaces. As regards the
technology capabilities, in producing complex aspherical surfaces with a
diameter of the order of several meters, an accuracy that ensures the
diffraction-limited image quality has been achieved in recent years.

In this paper, we represent Schwarzschild's formulas in a form that is valid
for arbitrary two-mirror aplanatic systems. Subsequently, we analyze the image
quality in aplanats calculated by extending the class of surfaces in the
standard ZEMAX optical program \footnote{ZEMAX Development Corporation, USA.}.
We show that exact Schwarzschild aplanats with a fast focal ratio provide much
better images than Ritchey--Chr\'etien and Gregory--Maksutov systems, which are
aplanatic only in the third order of the aberration theory.

\section*{FIRST-ORDER CHARACTERISTICS}

Depending on whether the secondary mirror precedes or succeeds the focus of the
primary mirror, the two-mirror telescopes are divided into two major classes:
\emph{prefocal} and \emph{postfocal} systems (Maksutov~1932, 1946;
Michelson~1976). The equivalent focal length~$F$ is positive for prefocal
systems, and is negative for postfocal systems. Each of the major classes
includes focal \emph{reducing} and \emph{lengthening} systems; $F$ is smaller
and larger in absolute value than the focal length of the primary mirror~$F_1$
for the former and the latter, respectively (we assume that ${F_1>0}$).

% ------------------------------------------------ Fig.01
\begin{figure}[t]
   \centering
   \includegraphics[width=0.50\textwidth]{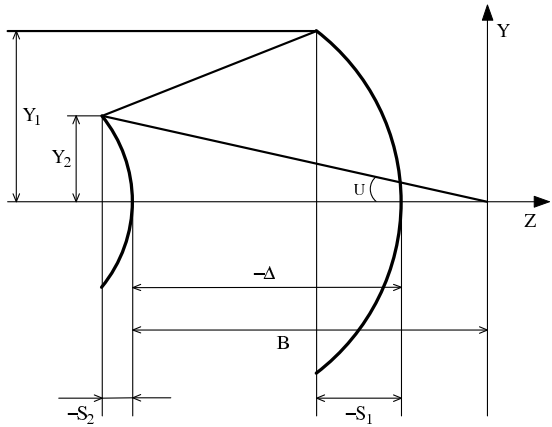}
   \caption*{Basic notation for a Cassegrain system as an example.}
   \label{f01}
\end{figure}
% -------------------------------------------- End Fig.01

Let $\Delta$ be the difference between the axial coordinates of the secondary
and primary mirror vertices; $B$ be the back focal length of the system; $Y_1$
be the height of the ray incident on the primary mirror parallel to the optical
axis; $Y_2$ be the height of this ray on the secondary mirror; $S_1$ and $S_2$
are the sags of the primary and secondary mirrors corresponding to this ray;
and~$U$ is the angle between the ray emergent from the system and its optical
axis. The notation used for a Cassegrain system is explained in Fig.~1. We
assume that the entrance pupil of the telescope is located on the primary
mirror, $\Delta$ is always negative, $Y_1$ and~$B$ are positive, the signs
of~$S_1$ and~$S_2$ are identical to those of the vertex radii of
curvature~$R_1$ and~$R_2$ for the corresponding mirrors, $Y_2$ and~$U$ are
positive for a prefocal system and negative for a postfocal system. Since the
system is axially symmetric, to describe the shape of the surfaces, it will
suffice to consider only their meridional section by the $(Y,Z)$ plane.

We choose the following four parameters as the specified characteristics of the
system:
\begin{itemize}
 \item[--] the diameter of the primary mirror~$D$,
 \item[--] the equivalent focal length of the telescope~$F$,
 \item[--] the relative value of the back focal length ${q \equiv B/F}$,
 \item[--] the quantity ${\beta \equiv F_1/F}$ that is the reciprocal of the
 magnification on the secondary mirror.
\end{itemize}
To avoid misunderstandings, the classification of two-mirror telescopes used
here is summarized in the Table~1. The parameters~$q$ and~$\beta$ can take on
any values from~$-\infty$ to~$\infty$. For slow telescopes, $|q|$ is
approximately equal to the linear central obscuration coefficient by the
secondary mirror. We are mainly interested in values of ${|q| < 1}$: it is in
this range the promising telescopes should be sought.

% ----------------------------------------------------------- Table01
\begin{center}
\begin{tabular}{cl|c|c}
 \multicolumn{4}{l}{\textit{\small{Table~1. Types of two-mirror telescopes}}}\\
 \hline
 \multicolumn{2}{c|}{System} & Lenthening & Shortening \\
 \hline
  Prefocal &  & Cassegrain & Schwarzschild \\
   $F, q, \beta > 0$     &  & $0<\beta<1$ & $\beta>1$ \\
 \hline
 Postfocal &  & \multicolumn{2}{c}{Gregory} \\
   $F, q, \beta < 0$     &  & $0<|\beta|<1$ & $|\beta|>1$ \\
 \hline
\end{tabular}
\end{center}
% ---------------------------------------------------------- End Table01

Evidently, the focal ratio of the telescope is ${\phi \equiv |F|/D}$, and the
remaining first-order characteristics of all the systems under consideration
may be represented in terms of the initial parameters as

$$
  \Delta = -(1-q)\beta F, \qquad R_1 = -2\beta F, \qquad R_2 =
  -\frac{2q\beta}{1-\beta}\,F.
  \eqno(1)
$$
For the subsequent analysis, it is convenient to introduce the special
designation

$$
  \delta \equiv (1-q)\beta,
  \eqno(2)
$$
so ${\Delta = -\delta F}$.

\section*{SCHWARZSCHILD's FORMULAS}

A necessary condition for the absence of coma is the satisfaction of the Abbe
\emph{sine condition} in the system,

$$
  Y_1/ \sin U = F = \mbox{\rm{const}},
  \eqno(3)
$$
for all of the ray heights~$Y_1$ and aperture angles~$U$. Schwarzschild managed
to rigorously combine the sine condition and the equation for the ray passage
in the system. The original calculations apply only to a reducing prefocal
system for ${\delta \neq 1}$, but they are so general in nature that can also
be performed for systems of other types with minor modifications. Since these
calculations are very cumbersome, we will not present them here, but only write
out the final formulas in a form that is valid for an arbitrary two-mirror
system.

The sought relations can be written in parametric form; they relate the
ordinates of the light ray ($Y_1$,~$Y_2$) at the points of its intersection
with the mirror surfaces and the mirror sags ($S_1$,~$S_2$) corresponding to
these points to the free parameter

$$
  t = \sin^2 (U/2), \qquad 0 \le t \le t_{\max}.
  \eqno(4)
$$
Equation~(3) and the condition ${|Y_1| \leq D/2}$ yield the maximum aperture
angle:
$$
  U_{\max} = \arcsin \left( 2\phi \right)^{-1},
  \eqno(5)
$$
and substituting this value into~(4) yields

$$
  t_{\max} = \frac{1}{2} \left[1-\sqrt{1-(2\phi)^{-2}}\, \right].
  \eqno(6)
$$
Usually, $t_{\rm{max}}$ is small; even for a fast system with a focal ratio of
${\phi = 1}$ we have ${t_{\rm{max}} \simeq 0.067}$.

Let us also define the index

$$
  \gamma = (1-\delta)^{-1}, \qquad  \delta \ne 1.
  \eqno(7)
$$
Then, the formulas that specify the primary mirror profile can then be written
as

$$
  Y_1/|F| = 2\sqrt{t(1-t)},
  \eqno(8')
$$
$$
  S_1/F = \left\{
  \begin{array}{ll}
  q \left[ 1-(1-t)^\gamma |1-t/\delta|^{2-\gamma} \right]
  - t(1-t)/\delta, \qquad  \delta \ne 1, &\\
  &\\
  q \left[ 1-(1-t)^2\,e^{t/(1-t)} \right]-t(1-t),
  \qquad  \delta = 1. &
  \end{array}
  \right.
  \eqno(8'')
$$
The analogous relations for the secondary mirror are

$$
  Y_2/|F| = 2\sqrt{t(1-t)}\,/\Theta(t),
  \eqno(9')
$$
$$
  S_2/F = q - (1-2t)\,/\Theta(t),
  \eqno(9'')
$$
where the auxiliary function

$$
  \Theta(t) = \left\{
  \begin{array}{ll}
  t/\delta + q^{-1}(1-t)^{1-\gamma} |1-t/\delta|^\gamma,
  \qquad \delta \ne 1, &\\
  &\\
  t+q^{-1}\,(1-t)\,e^{-t/(1-t)}, \qquad \delta = 1. &
  \end{array}
  \right.
  \eqno(10)
$$

Note that the results for ${\delta = 1}$ can be obtained both by repeating
calculations similar to Schwarzschild's original calculations and by passing to
the limit ${\delta \rightarrow 1}$. According to~(2), ${\beta = (1-q)^{-1}}$ in
the case under consideration, and formulas~(1) take the form

$$
  \Delta = -F, \qquad R_1 = -\frac{2}{1-q}\,F, \qquad R_2 = 2F.
  \eqno(1')
$$

Thus, assigning values from the interval $[0,t_{\rm{max}}]$ to the free
parameter~$t$, we find the profile of the mirror surfaces from
formulas~(8)--(10) that ensure the absence of spherical aberration and coma in
an arbitrary two-mirror telescope.

\section*{SLOW SYSTEMS}

According to~(6), the upper boundary of the free parameter~$t_{\rm{max}}$ is
less than~$0.01$ at ${\phi = 2.6}$ and, at an even slower focal ratio, is

$$
  t_{\max} \simeq (4 \phi)^{-2} \ll 1.
  \eqno(11)
$$
For this case, Schwarzschild~(1905) expanded the exact formulas in power series
of the normalized ray heights ${y_1 \equiv Y_1/F}$ and $y_2 \equiv Y_2/F$, the
first terms of which in our notation are

$$
  S_1/F = -\frac{1-q}{4\delta}\,y_1^2 + \frac{q}{32\delta}\,y_1^4
  + q\,\frac{1+4\delta}{384\delta^2}\,y_1^6 + q\,
  \frac{2+11\delta+30\delta^2}{6144\delta^3}\,y_1^8 + \ldots\,,
  \eqno(12)
$$
$$
  S_2/F = -\frac{1-q-\delta}{4q\delta}\,y_2^2 + \frac{2-\delta-
  4q+2q^2+2q\delta}{32q^3\delta^2}\,y_2^4 + \ldots \,.
  \eqno(13)
$$
The expansions are also valid for ${\delta = 1}$.

The sag of an arbitrary conic section~$S_{\rm{c}}$ is known to be

$$
  S_{\rm{c}}/F = \frac{y^2}{r}\,\left[ 1+\sqrt{1-
  (1-\epsilon^2)(y/r)^2}\, \right]^{-1},
   \eqno(14)
$$
where ${r = R/F}$ is the dimensionless radius of curvature at the vertex,
and~$\epsilon$ is the surface eccentricity. Hence, one can easily obtain the
expansion

$$
  S_{\rm{c}}/F = \frac{1}{2r}\,y^2 +\frac{1-\epsilon^2}{8r^3}\,y^4 +
  \frac{(1-\epsilon^2)^2}{16r^5}\,y^6 + \ldots \,.
  \eqno(15)
$$
Its comparison with~(12) and~(13) allows the approximation of exact aplanatic
surfaces by conicoids for slow systems to be elucidated.

For the primary mirror, the first two expansion terms give the following
expressions for the radius of curvature and the square of the eccentricity:

$$
  r_1 = -\frac{2\delta}{1-q}\,, \qquad
  \epsilon_1^2 = 1 + \frac{2q\delta^2}{(1-q)^3}\,.
  \eqno(16)
$$
Give the definition of the constant~$\delta$ in~(2), we see that the former
expression is identical to that in~(1), while it follows from the latter
expression that

$$
  \epsilon_1^2 = 1+\frac{2q\beta^2}{1-q}\,.
  \eqno(17)
$$
Just as above, we obtain for the secondary mirror:

$$
  r_2 = -\frac{2q\delta}{1-q-\delta}\,, \qquad
  \epsilon^2_2 = \left( \frac{1-q+\delta}{1-q-\delta} \right)^2 +
  \frac{2\delta^2}{(1-q-\delta)^3}\,.
  \eqno(18)
$$
The expression for~$r_2$ is identical to that in~(1), while the square of the
eccentricity of the secondary mirror expressed in terms of the basic variables
$(q,\beta)$ can be written as

$$
  \epsilon_2^2 = \left( \frac{1+\beta}{1-\beta} \right)^2 +
  \frac{2\beta^2}{(1-q)(1-\beta)^3}\,.
  \eqno(19)
$$

Equations~(17) and~(19) form the basis for the theory of systems aplanatic in
the third order of the aberration theory (see, e.g., Schroeder~2000, \S6.2.b).
For clarity, Fig.~2 gives a graphical representation of these expressions that
is essentially identical to that in the book by Maksutov~(1946, \S21). An
afocal Mersenne system of two paraboloids with coincident foci corresponds to
${\beta = 0}$. For ${0 < \beta < 1}$, Eqs.~(17) and~(19) describe a
Ritchey--Chr\'etien system. The case of ${\beta = 1}$ corresponds to the
so-called \emph{ring} telescope with a planoid secondary mirror (Maksutov~1932;
Churilovskii~1958; Landi Dessy~1968). The domain ${\beta> 1}$ corresponds to
the family of reducing Schwarzschild telescopes aplanatic in the third-order
approximation. In the entire domain ${\beta> 0}$, the primary mirror is a
hyperboloid, while the secondary mirror is a hyperboloid in a lengthening
aplanat and changes its shape for reducing systems from an oblate
ellipsoid\footnote{This is the region of the ellipsoidal surface around the
point of its intersection with the minor axis; previously, an oblate ellipsoid
was commonly called an oblate spheroid.} to a hyperboloid, passing through a
sphere and a prolate ellipsoid. The domain ${\beta < 0}$ corresponds to the
aplanatic version of the Gregory system. Here, the secondary mirror remains a
prolate ellipsoid, while the primary mirror changes its shape from a prolate to
an oblate ellipsoid, also passing through a sphere.

% ------------------------------------------------ Fig.02
\begin{figure}[t]
   \centering
   \includegraphics[width=0.80\textwidth]{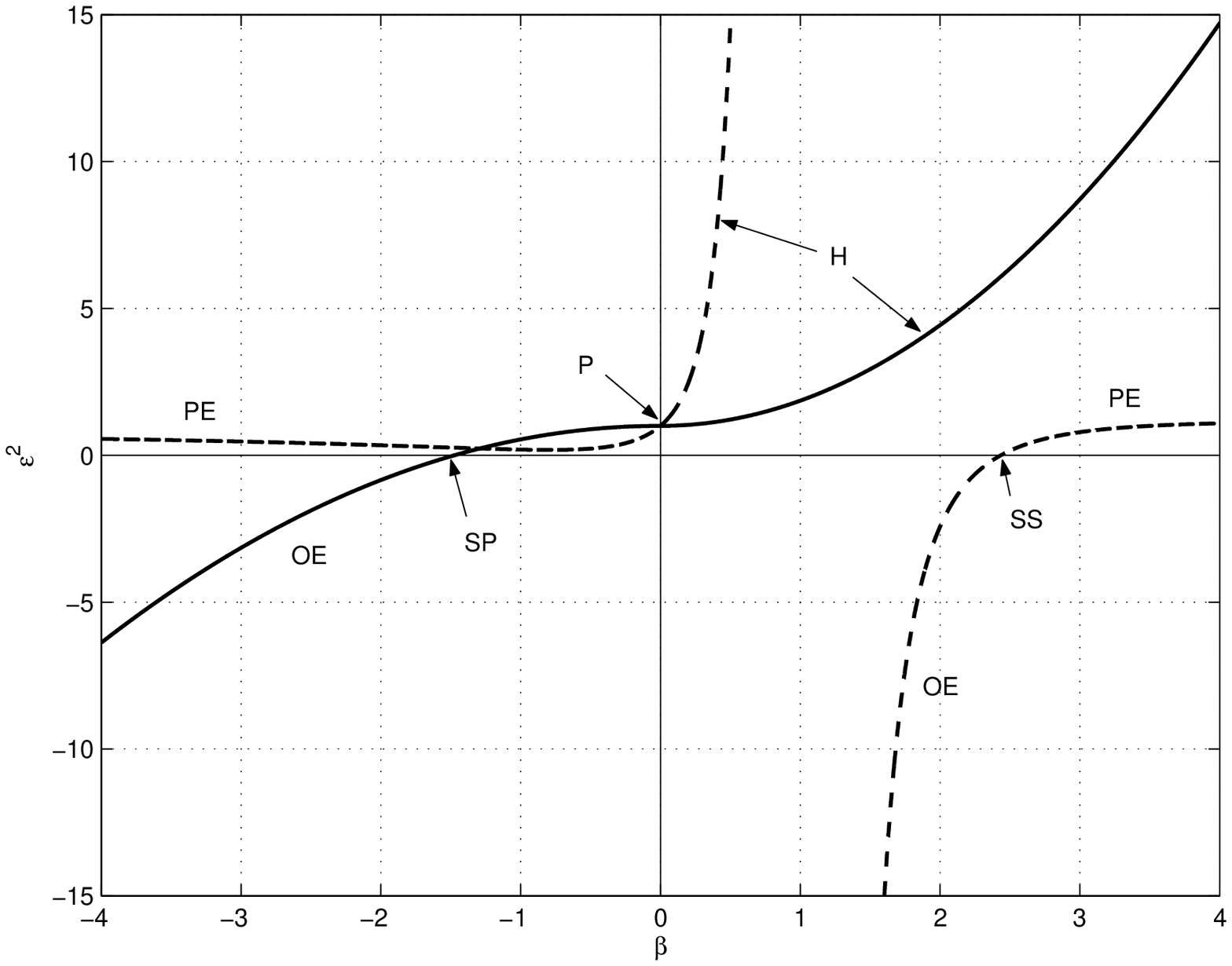}
   \caption*{Squares of the eccentricities of the primary (solid line) and
secondary (dashed line) mirrors versus~$\beta$ in approximate aplanats at ${|q|
= 0.30}$:\\ P~-- paraboloid, H~-- hyperboloid, PE~-- prolate ellipsoid, OE~--
oblate ellipsoid, SP~-- spherical primary mirror, SS~-- spherical secondary
mirror.}
   \label{f02}
\end{figure}
% -------------------------------------------- End Fig.02

We emphasize that Fig.~2 describes slow systems in which only the third-order
aberrations were corrected. Therefore, it gives only a rough idea of the family
of exact aplanatic surfaces.

\section*{SCHWARZSCHILD SURFACES IN THE\\ ZEMAX PROGRAM}

Ascertaining the image quality in exact Schwarzschild aplanats requires either
developing a special program for calculating the ray path in such systems or
extending the class of surfaces in one of the existing optical programs. We
chose the second way, especially since the powerful ZEMAX optical program
allows it to be implemented with relative ease. This requires writing
additional programs in the C/C${++}$ language in which the new surfaces
(8)--(10) and the optics based on them are described and then compiling these
programs and dynamically linking them with the main program. Thus, we can use
the extensive set of tools for studying the properties of optical systems
provided by ZEMAX.

In this way, we created the files \textit{ksp.dll} and \textit{kss.dll} (from
\textbf{K}arl \textbf{S}chwarzschild \textbf{p}rimary/\textbf{s}econdary) that
define, respectively, the primary and secondary mirrors of an arbitrary
aplanat. The quantities $(D,F,q,\beta)$ defined in the section entitled
``First-Order Characteristics'' should be specified as additional parameters
when calling a surface.

\section*{EXAMPLES OF FAST APLANATS}

An analysis of several model systems indicates that Ritchey--Chr\'etien and
Gregory--Maksutov systems yield a satisfactory approximation to exact aplanatic
systems for~$\phi$ of~$\sim4$ or larger. Volume of this publication does not
allow us to illustrate the change in image quality as the focal ratio of the
system becomes faster; here, we restrict our analysis to several examples. The
optical telescope layouts considered below should give a general idea of the
quality of the images obtained with exact aplanats.

We characterize the image quality by the quantity~$D_{80}$, the diameter of the
circle within which 80\% of the energy is concentrated in the diffraction image
of a point source. Denote the diameter of the field of view in which~$D_{80}$
does not exceed one arcsecond by~$2w_{\rm{oas}}$.

The astigmatism and the field curvature remain uncorrected in aplanatic
systems. Usually, these aberrations in mirror telescopes are removed by
inserting lens correctors. A preliminary analysis shows that a simple corrector
of three lenses with spherical surfaces inserted in a lengthening $f/1.2$
Schwarzschild aplanat provides flat field of view about~$1^\circ$ in diameter
(see section ``Including Schwarzschild Aplanats in Complex Systems'' below). We
are going to systematically study the correctors in Schwarzschild aplanats
later.

\subsection*{The Cassegrain System}

Figure~3 shows the optical layout of the SA-01 telescope with an aperture
diameter of 1~m and a focal ratio of ${\phi = 1.2}$. We chose ${q=0.30}$ and
${\beta=0.65}$ for the remaining parameters, so the magnification on the
secondary mirror is $m \equiv 1/\beta \simeq 1.54$. The focal ratio for the
primary mirror is $\phi_1 = 0.78$, which is achievable at the present state of
the art.

% ------------------------------------------------ Fig.03
\begin{figure}[t]
   \centering
   \includegraphics[width=0.50\textwidth]{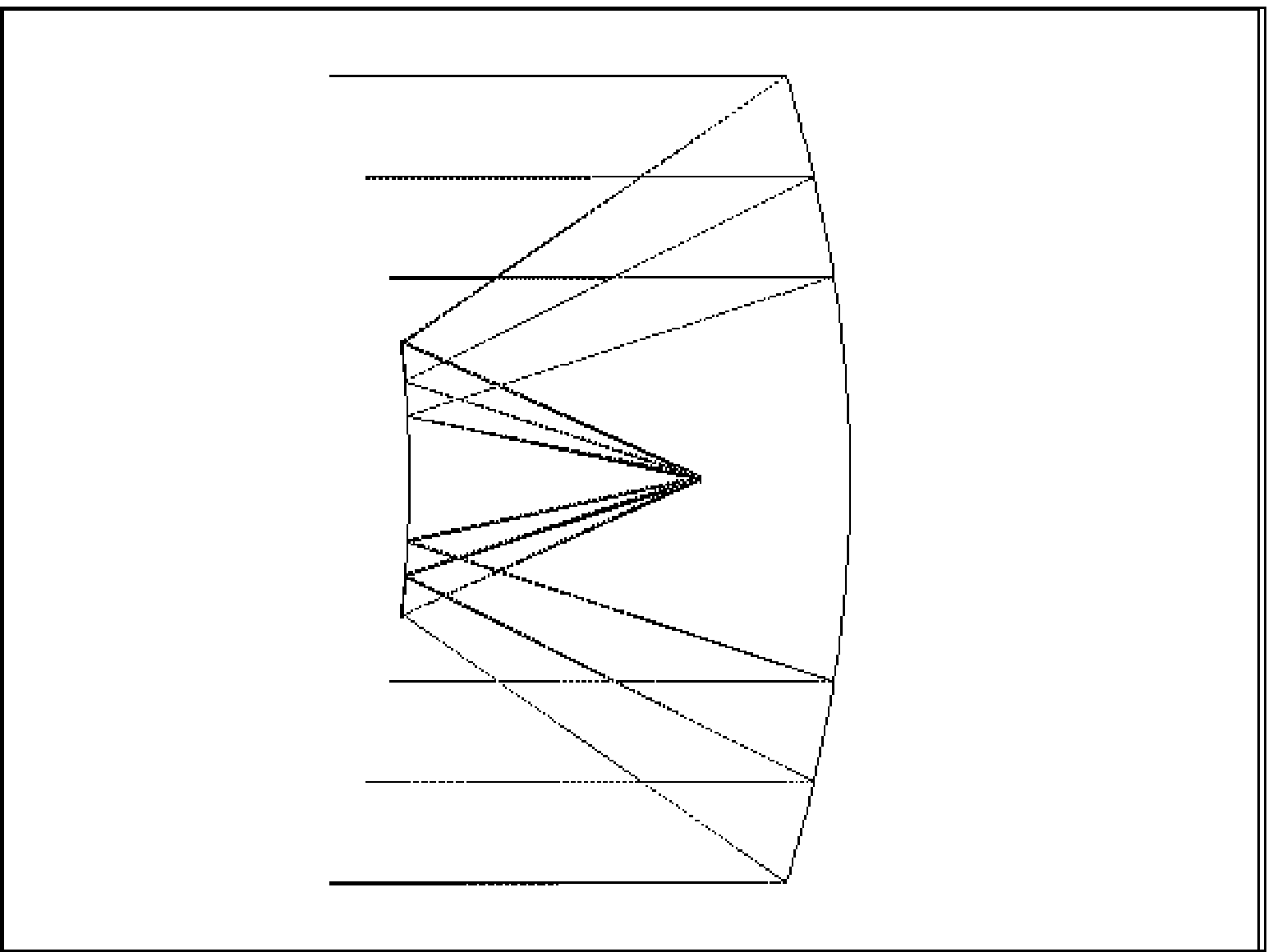}
   \caption*{Optical layout of the SA\mbox{-01} telescope with a focal ratio
    of $\phi = 1.2$.}
   \label{f03}
\end{figure}
% -------------------------------------------- End Fig.03
% ------------------------------------------------ Fig.04
\begin{figure}[t]
   \centering
   \includegraphics[width=0.60\textwidth]{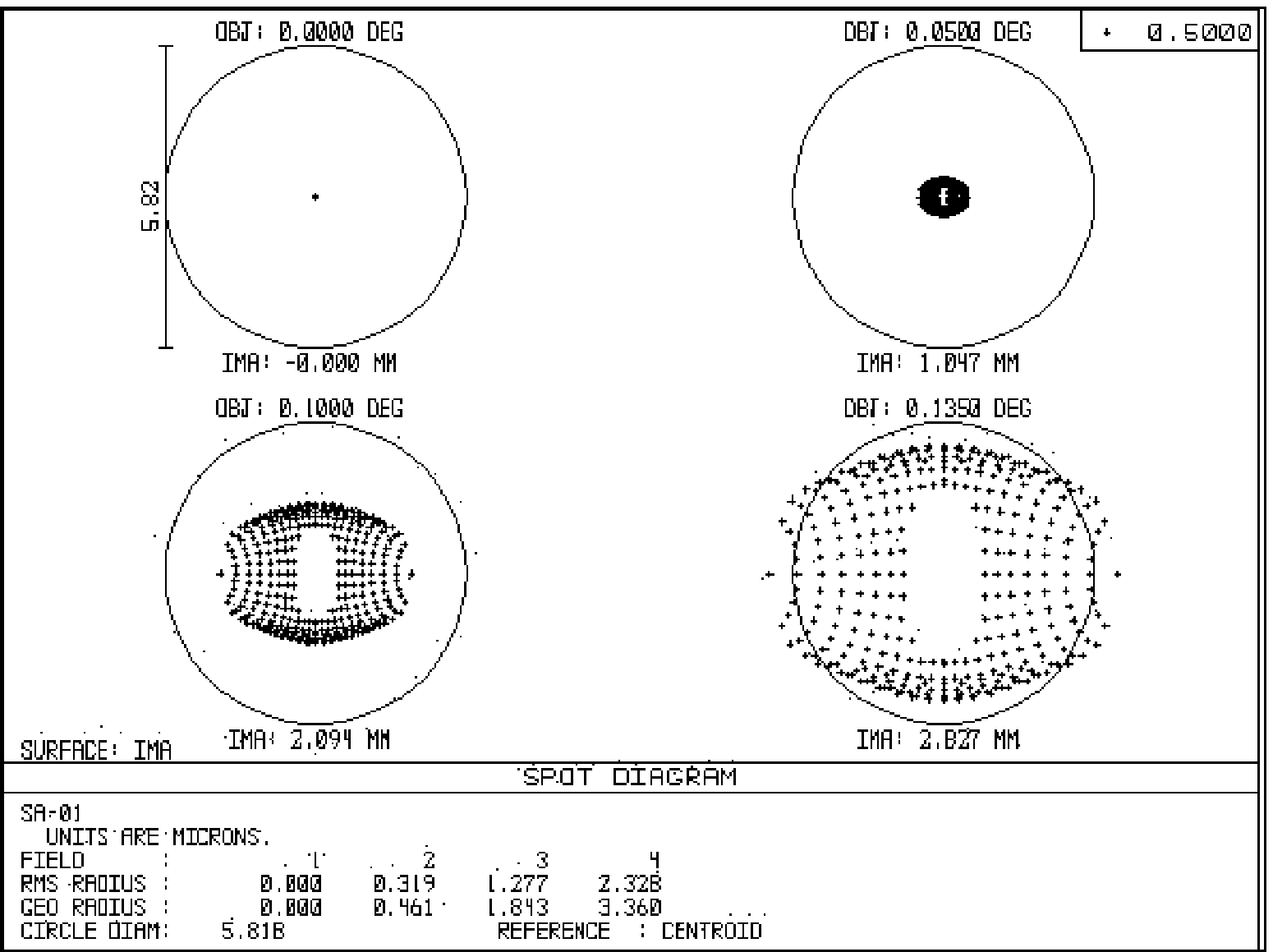}
   \caption*{Spot diagrams for the SA\mbox{-01} telescope in the field of view
$2w=0^\circ.27$\\ (the wavelength is 0.5~$\mu$m). The root-mean-square (RMS)
and geometrical (GEO) image radii as well as the diameter of the circle
corresponding to~$1''$ are given in micrometers.}
   \label{f04}
\end{figure}
% -------------------------------------------- End Fig.04
% ------------------------------------------------ Fig.05
\begin{figure}[t]
   \centering
   \includegraphics[width=0.50\textwidth]{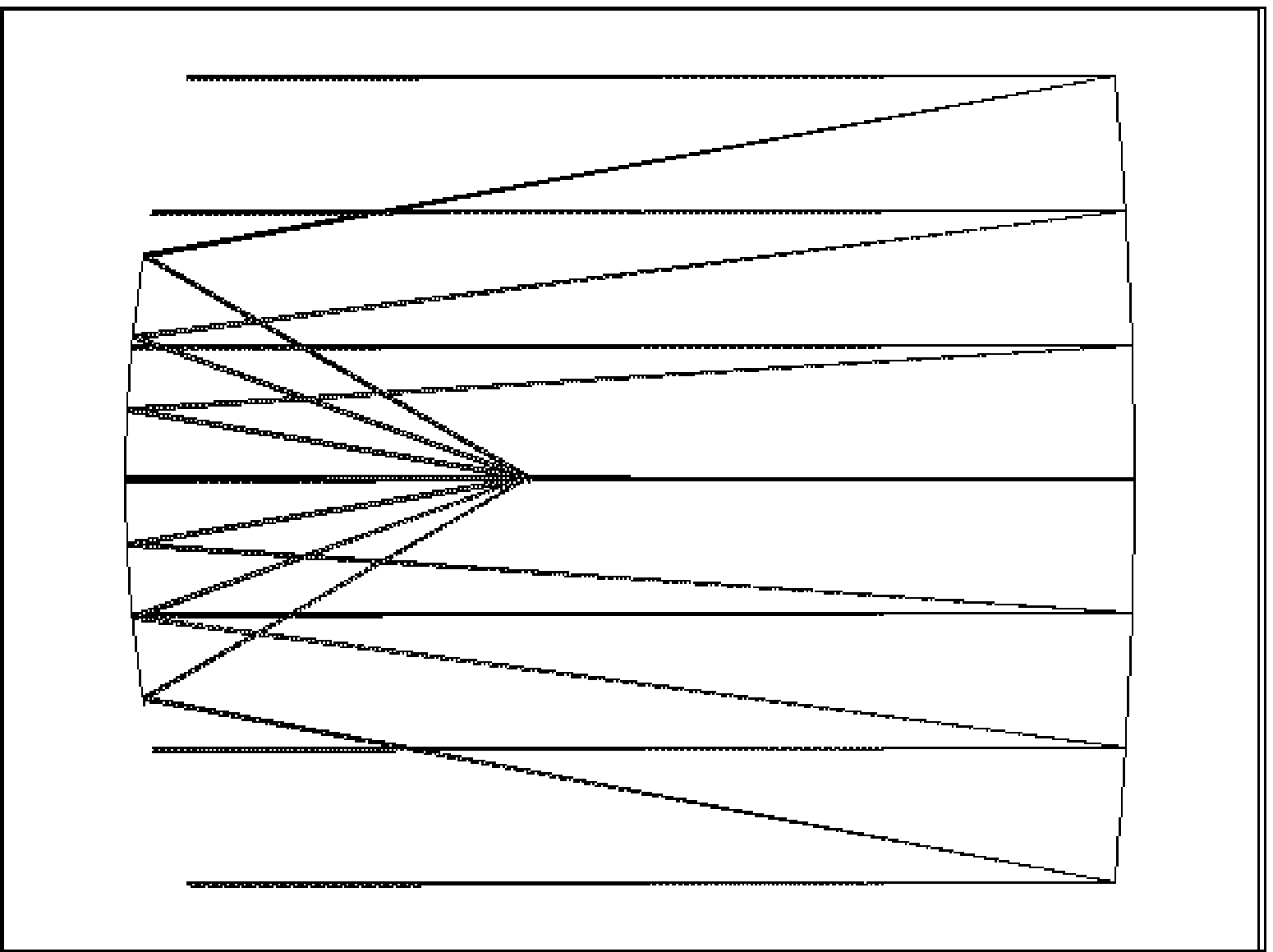}
   \caption*{Optical layout of the SA\mbox{-02} telescope with
    a focal ratio of $\phi = 1.0$.}
   \label{f05}
\end{figure}
% -------------------------------------------- End Fig.05
% ------------------------------------------------ Fig.06
\begin{figure}[t]
   \centering
   \includegraphics[width=0.60\textwidth]{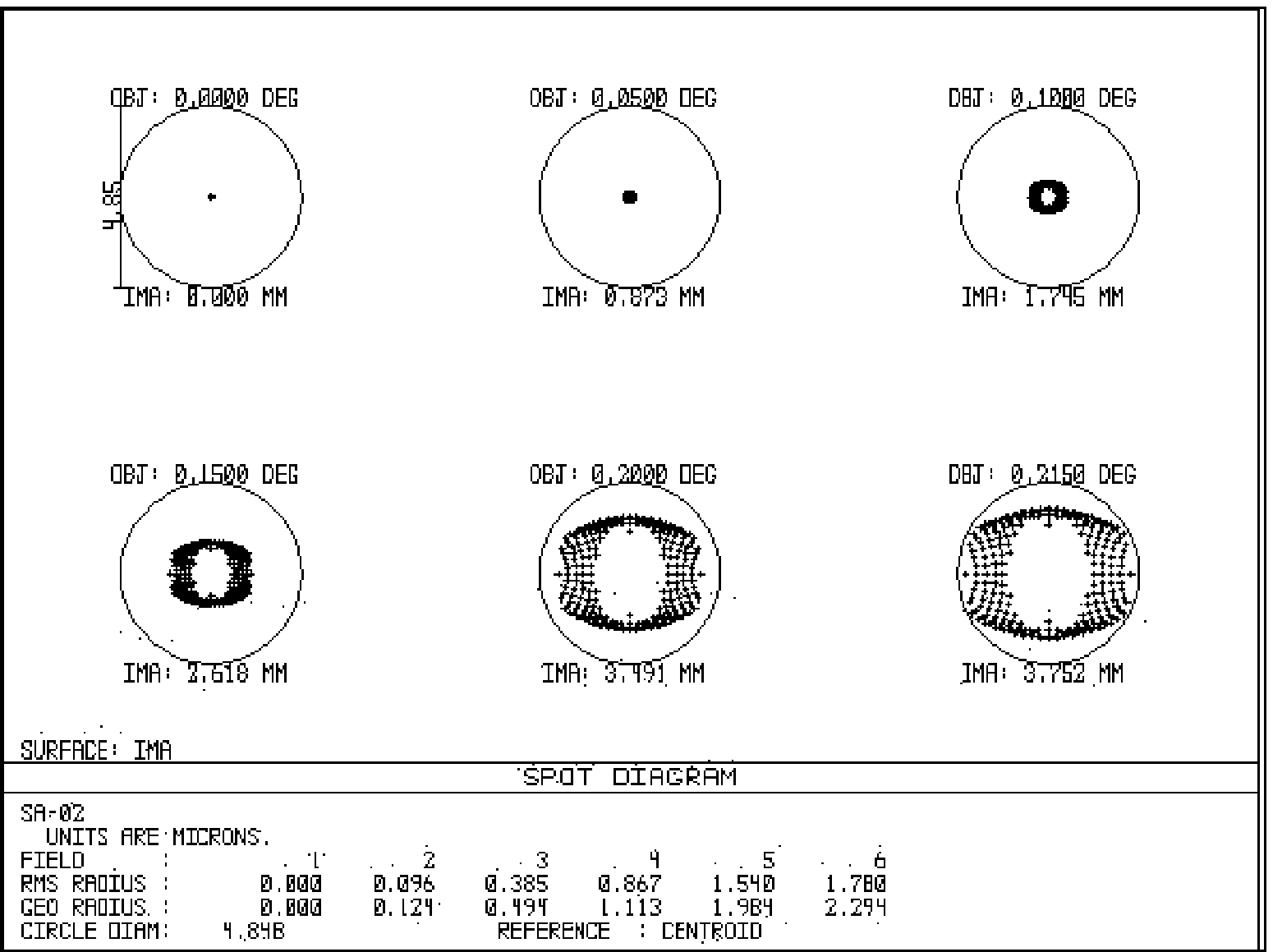}
   \caption*{Spot diagrams for the SA\mbox{-02} telescope in the field of view
$2w=0^\circ.43$\\ (the wavelength is 0.5~$\mu$m). The RMS and geometrical (GEO)
image radii as well as the diameter of the circle corresponding to~$1''$ are
given in micrometers.}
   \label{f06}
\end{figure}
% -------------------------------------------- End Fig.06
% ------------------------------------------------ Fig.07
\begin{figure}[t]
   \centering
   \includegraphics[width=0.60\textwidth]{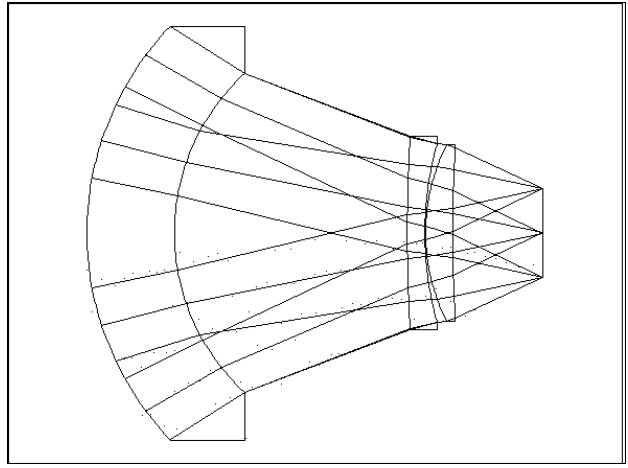}
   \caption*{Three-lens corrector to the SA\mbox{-}01 telescope.}
   \label{f07}
\end{figure}
% -------------------------------------------- End Fig.07
% ------------------------------------------------ Fig.08
\begin{figure}[t]
   \centering
   \includegraphics[width=0.75\textwidth]{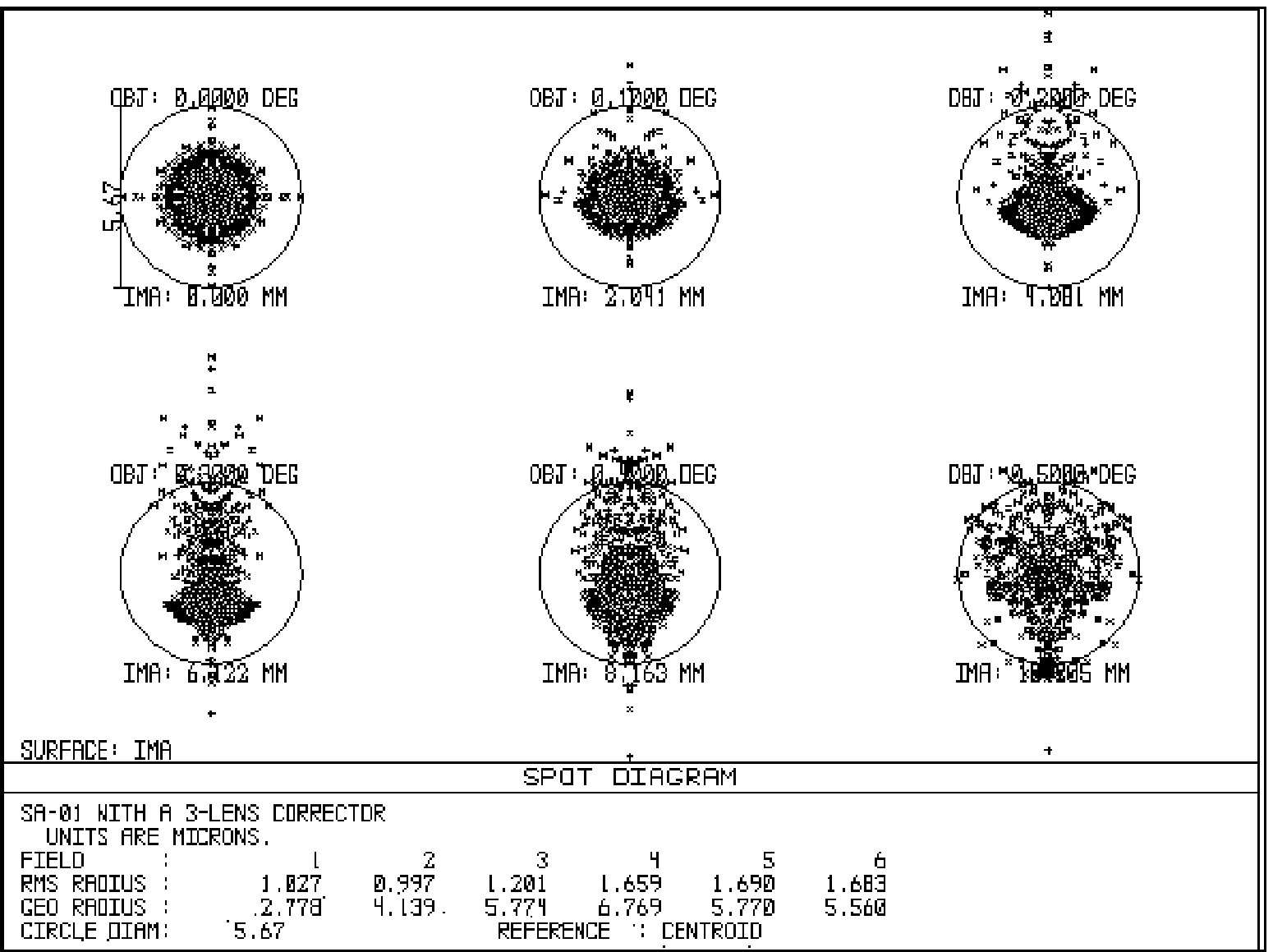}
   \caption*{Spot diagrams for the SA\mbox{-01} telescope with the
corrector shown in Fig.~7. The diameter of the flat field of view is
$2w=1^{\circ}$; polychromatic light with wavelengths of 0.4--1.0~$\mu$m is
considered. The RMS and geometrical (GEO) image radii as well as the diameter
of the circle corresponding to~$1''$ are given in micrometers.}
   \label{f06}
\end{figure}
% -------------------------------------------- End Fig.08

The field of view~$2w_{\rm{oas}}$ is $0^\circ.27$ ($16'.2$) in this case; the
image quality is illustrated by the spot diagrams\footnote{The \textit{spot
diagram} is the distribution of light rays in the image of a star on the focal
surface.} in Fig.~4. The Airy circle is 1.5~$\mu$m in diameter, so the image
quality in the $10'$ central area is determined by the diffraction of light.
Given the field of view, the central obscuration coefficient is 0.34. The
optimum radius of curvature of the focal surface is $R_{\rm{im}} = -281.4$~mm,
so maximum deviation of this surface from the plane in the field of view is
14.2~$\mu$m. Since the system is fast, direct imaging with a flat detector
would increase the root-mean-square (RMS) image radius at the edge of the field
from~2.3~$\mu$m to 5.2~$\mu$m. The system is compact: its length, 546~mm, is
almost half the aperture diameter.

In the Ritchey--Chr\'etien system, RC-01, with the same initial parameters, the
squares of the mirror eccentricities are $\epsilon_1^2=1.362143$ and
$\epsilon_2^2=50.379425$. The RMS image diameter of a point source on the
optical axis is more than $0.5~\mbox{mm}$, so it would be unreasonable to talk
about a useful field of view.

For the coincident vertices of the primary mirrors in the exact aplanat, SA-01,
and the Ritchey--Chr\'etien system, RC-01, the difference between their
profiles changes monotonically, reaching 88.9~$\mu$m at the edge; the sag for
SA\mbox{-}01 is smaller in absolute value than that for RC\mbox{-01}. A similar
picture is also observed for the secondary mirrors: the largest deviation of
the surfaces is 180.6~$\mu$m, and the sag for SA-01 is smaller in absolute
value than that for RC\mbox{-01}.

\subsection*{The Schwarzschild System}

As an example, Schwarzschild~(1905) considered a reducing telescope with a
diameter of 1~m and a focal ratio of ${\phi = 1.0}$; recalculating the original
parameters to those used here yields $q = 0.50$ and $\beta = 2.50$. The focal
ratio for the primary mirror is $\phi_1 = 2.50$. Figures~5 and~6 are similar to
Figs.~3 and~4, respectively.

In this case, the field of view $2w_{\rm{oas}}$ is $0^\circ.43$. Given the
field of view, the central obscuration coefficient is 0.56. The optimum focal
surface is close to a plane, $R_{\rm{im}} = 2758$~mm, so imaging with a flat
detector at such a fast focal ratio of the system would increase the RMS image
radius at the edge of the field from~1.8~$\mu$m to only 2.1~$\mu$m. The length
of the system is 1250~mm.

The RMS image diameter on the optical axis in the Ritchey--Chr\'etien system
with the same initial parameters reaches almost 2~mm (the back focal length and
the radius of curvature of the focal surface were optimized).

\subsection*{Including Schwarzschild Aplanats in Complex Systems}

It is quite clear that Schwarzschild aplanats are not only of interest in their
own right, but can also form the basis for complex optical systems: telescopes,
fast objectives, etc. As an example, we restrict ourselves to a three-lens
corrector to the SA\mbox{-01} system considered above. The aim is to make the
field of view flat and to increase the diameter of the field of subarcsecond
image quality from the initial $0^\circ.27$ to about~$1^\circ$, while retaining
the focal ratio of the system at the previous level of ${\phi \simeq 1.2}$. For
simplicity, we assume that all of the lenses surfaces are spherical in shape
and that the parameters of the two-mirror telescope are fixed.

The optical layout of this corrector is shown in Fig.~7; the corresponding spot
diagrams in polychromatic light with wavelengths from~0.40~$\mu$m to 1.0~$\mu$m
are shown in Fig.~8. The focal ratio of the system is $\phi = 1.169$. The lens
diameters are small, they are equal to (in the ray path) 94.7, 44.1,
and~40.2~mm; the types of glass are Schott SF15, LAK16A, and SK11,
respectively; the back focal length (the gap between the last lens and the
detector) is $20~\mbox{mm}$. The image diameter of a point source~$D_{80}$ at
the edge of a $1^\circ$ field of view is $1''.02$. Naturally, the image quality
is better in the narrow spectral bands where observations are commonly
performed. A larger field of view can be achieved by simultaneously optimizing
the correctors and the two mirrors.

\section*{CONCLUDING REMARKS}

First of all, it is necessary to note that a field of~${\sim1^\circ}$ in
diameter is insufficient to carry out the sky surveys mentioned in the
Introduction. Either a single mirror with a lens corrector at the prime focus
(Blanco \emph{et al.}~2002; McPherson \emph{et al.}~2002; Terebizh~2004) or an
aplanatic (in the third order) two-mirror system supplemented with a corrector
(Angel \emph{et al.}~2000; York \emph{et al.}~2000; Hodapp \emph{et al.}~2003)
forms the basis for the currently designed telescopes with a ~${2^\circ -
3^\circ}$ field of view. A lens corrector at the prime focus of a hyperbolic
mirror allows a $3^\circ$ field of view of subarcsecond quality to be achieved
by relatively simple means: all lenses are made of glass of the same, virtually
arbitrary type, and their surfaces are spherical, but the sizes of the entire
system are large. On the other hand, the telescopes based on the fast
Cassegrain design are quite compact, but the lens or mirror--lens corrector
that supplements them is complex. The choice of a Schwarzschild aplanat as the
basic two-mirror system may allow the lens part of wide-field telescopes to be
simplified, while preserving their compactness.

In principle, the production of Schwarzschild surfaces poses no specific
problems compared to the production of conicoids. In both cases the
difficulties rapidly grow as the focal ratio becomes faster, so particular
attention should be paid to the reliability of optics control. Offner's
conventional system with two small lenses or mirrors has the number of degrees
of freedom that is large enough to achieve focal ratios ${\phi \sim 1}$
(Offner~1978). The system with a Hartmann diaphragm in the convergent beam
proposed by Biryukov and Terebizh~(2003) also seems promising; the relative
complexity of the data analysis is compensated by the exceptional ease of the
experimental facility.

The difficulty of producing the surfaces described by Eqs.~(8)--(10) can be
roughly estimated if we approximate them by conic sections. The largest
deviation of a conicoid from the nearest sphere is known to be

$$
  \delta S_{\max} = \frac{D\, \epsilon^2}{4096\, \phi^3}\,,
  \eqno(20)
$$
where $D$, $\epsilon$, and~$\phi$ are, respectively, the diameter,
eccentricity, and focal ratio of the surface (see, e.g., Maksutov~1946, \S20).
For example, the asphericity of the 1.8-m Vatican Observatory telescope primary
mirror ($\phi=1.0$, $\epsilon^2=1.0$) is very large, 440~$\mu$m. Nevertheless,
in the early~1990s, this mirror was made with an accuracy that ensured the RMS
error of the wavefront\footnote{Recall that, according to Mar\'echal's known
criterion, an optical system may be considered to be diffraction-limited if the
RMS error of the wavefront does not exceed $\sim1/14$ of the wavelength (Born
and Wolf~1999, \S9.3).} equal to only $1/30$ of the wavelength
$\lambda=0.5$~$\mu$m (Update on Progress at the Steward Obs.~1991).

Dierickx~(1999) noted that the \emph{relative slope} of the aspherical surface
rather than the departure from the nearest sphere is important. It is
convenient to use the following dimensionless quantity as a parameter that
characterizes the difficulty of producing a surface:

$$
  \omega = \frac{\epsilon^2}{8\, \phi^3}\,,
  \eqno(21)
$$
the reciprocal of the parameter~$\delta c$ introduced by Dierickx. The point is
that~$\omega$ is proportional to the relative slope of the aspherical surface.
Thus, a greater production difficulty corresponds to a larger~$\omega$, and
characteristics~(20) and~(21) become similar in this sense. For the mirror of
the Vatican Observatory mentioned above, ${\omega = 1/8}$. At present, fast
mirrors with $\epsilon^2 \simeq 5-10$ are produced, so we may consider ${\omega
\sim 1}$ to be achievable.

Let us consider the SA\mbox{-01} system from this point of view. For the
primary mirror, $\epsilon_1^2 \simeq 1.36$ and $\phi_1 = 0.78$, which yields
$\omega_1 = 0.36$. For the secondary mirror, $\epsilon_2^2 \simeq 50.4$ and
$\phi_2 \simeq 2$, which yields $\omega_2 \simeq 0.79$. We see that producing
aplanatic surfaces is a serious, but resolvable problem.

The question of light baffles is particularly acute for fast systems. Recently,
exact analytical expressions have been derived for the sizes of \emph{optimum}
baffles in a Cassegrain system with mirrors in the shape of arbitrary conic
surfaces (Terebizh~2001). Similar expressions can also be easily derived for
postfocal systems. Since Schwarzschild surfaces differ from conicoids, the
mentioned result is generally inapplicable to exact aplanats, but the path of
rays in them does not change so significantly as do the images, and the
analytical expressions can be used to obtain a satisfactory approximation.

Having in mind aplanatic (in the third order) systems, Maksutov~(1946) wrote:
``Aplanatic mirror telescopes have a great future, since only these systems
will allow superpowerful and fast instruments with the fields of first-rate
photographic images sufficient for practical purposes to be made.'' This
opinion seems to be all the more valid for exact Schwarzschild aplanats.

\section*{ACKNOWLEDGMENTS}

I am grateful to V.~V.~Biryukov (Moscow State University) for a helpful
discussion of the questions touched upon here.

\vspace{0.2cm} \hspace{5cm}
 Translated by V.~Astakhov

\end{document}